\documentclass[british,submission,Phys]{SciPost}
\usepackage[T1]{fontenc}
\usepackage[utf8]{inputenc}
\usepackage{geometry}
\usepackage{color}
\usepackage{mathrsfs}
\usepackage{amsmath}
\usepackage{amssymb}
\usepackage{graphicx}

\makeatletter
\usepackage[T1]{fontenc}
\usepackage[section]{placeins}

\hypersetup{
    colorlinks=true,       % false: boxed links; true: colored links
    linkcolor={red!50!black},        % color of internal links
    citecolor={blue!80!black},        % color of links to bibliography
    filecolor=magenta,     % color of file links
    urlcolor={blue!80!black}
}

\DeclareUnicodeCharacter{2009}{\,} 
\let\mySection\section\renewcommand{\section}{\suppressfloats[t]\mySection}

\makeatother

\usepackage{babel}
\begin{document}
\global\long\def\vect#1{\overrightarrow{\mathbf{#1}}}%

\global\long\def\abs#1{\left|#1\right|}%

\global\long\def\av#1{\left\langle #1\right\rangle }%

\global\long\def\ket#1{\left|#1\right\rangle }%

\global\long\def\bra#1{\left\langle #1\right|}%

\global\long\def\tensorproduct{\otimes}%

\global\long\def\braket#1#2{\left\langle #1\mid#2\right\rangle }%

\global\long\def\b#1{\boldsymbol{#1}}%

\global\long\def\bb#1{\mathbf{#1}}%

\global\long\def\cal#1{\mathcal{#1}}%

\global\long\def\scr#1{\mathscr{#1}}%

\global\long\def\braket#1#2{\left\langle #1|#2\right\rangle }%

\global\long\def\p#1{\left(#1\right)}%

\global\long\def\t#1{\text{#1}}%

\global\long\def\s#1{{\displaystyle {\displaystyle #1}}}%

\global\long\def\com#1#2{\left[#1,#2\right]}%

\global\long\def\an#1#2{\left\{  #1,#2\right\}  }%

\global\long\def\omv{\overrightarrow{\Omega}}%

\global\long\def\inf{\infty}%

\begin{center}{\color{blue!30!black}\Large \textbf{Unambiguous Simulation
of Diffusive Charge Transport in Disordered Nanoribbons\\ }\color{black}}\end{center}

\begin{center}

\textbf{H. P. Veiga\textsuperscript{1,2$\dagger$}, S. M. João\textsuperscript{1,3},
J. M. Alendouro Pinho\,\textsuperscript{1}, J. P. Santos Pires\,\textsuperscript{1},
and J. M. Viana Parente Lopes\textsuperscript{1,$\ddagger$}}

\end{center}

\begin{center}{\bf1} Centro de Física das Universidades do Minho
e do Porto (CF-UM-UP) and Laboratório de Física para Materiais e Tecnologias
Emergentes LaPMET, University of Porto, 4169-007 Porto, Portugal\\{\bf2}
Institute for Systems and Computer Engineering, Technology and Science
(INESC TEC), 4150-179 Porto, Portugal\\{\bf3} Department of Materials,
Imperial College London, South Kensington Campus, London SW7 2AZ,
United Kingdom\end{center}

\vspace{-0.50cm}\begin{center}\textsuperscript{$\dagger$} up201805202@edu.fc.up.pt,
{$\ddagger$} jlopes@fc.up.pt\end{center}

\begin{center} 
\today 
\end{center}

\section*{\vspace{-0.5cm}\color{blue!30!black}Abstract\color{black}\vspace{-0.2cm}}

{\bf Charge transport in disordered two-dimensional (2D) systems
showcases a myriad of unique phenomenologies that highlight different
aspects of the underlying quantum dynamics. Electrons in such systems
undergo a crossover from ballistic propagation to Anderson localization,
contingent on the system's effective coherence length. Between the
extended and localized phases lies a diffusive crossover in which
the charge conductivity is properly defined.

The numerical observation of this regime has remained elusive because
it requires fully coherent transport to be simulated in systems whose
dimensions are sufficiently large to meaningfully split the mean-free
path and localization length scales. To address this challenge, we
employed a novel linear scaling time-resolved approach that enabled
us to derive the dc-transport characteristics and observe the three
expected 2D transport regimes --- \emph{ballistic}, \emph{diffusive},
and \emph{localized}.}

\vspace{0.2cm}\hspace{-0.6cm}\color{blue!40!black}\rule{15.5cm}{0.04cm}
\vspace{-0.4cm}\color{black}

\tableofcontents{}\hspace{-0.6cm}\color{blue!40!black}\rule{15.5cm}{0.04cm}\color{black}

\section{\protect\label{sec:Introduction}Introduction}

As first understood by P. W. Anderson\textcolor{black}{$\,$\cite{anderson_absence_1958}},
the existence of quenched disorder can drastically influence how electrons
wander along an applied electric field in a metallic system. As it
gets stronger, disorder eventually causes all single-particle eigenstates
to exponentially localize in space, converting the system into a bulk
insulator where charge transport only occurs via the thermally activated
long-range hopping of electrons between distant localization centers\textcolor{black}{\,\cite{mott_conduction_1969}}.
While this is a generic behaviour of disordered lattices, the onset
of Anderson localization as a function of disorder strength is a process
that depends crucially on the system’s dimensionality. For one-dimensional
(1D) electrons, it is well known that an infinitesimally weak amount
of disorder will be strong enough to localize all quantum states\textcolor{black}{\,\cite{mott_theory_1961,borland_nature_1963}}
while, in three-dimensions (3D), a system is generally expected to
remain metallic up to a very strong disorder\textcolor{black}{\,\cite{paalanen_critical_1983,katsumoto_fine_1987,kramer_localization_1993,schubert_distribution_2010}}.
Just between these two situations lie the two-dimensional (2D) systems,
which formally define the lower critical dimension for the Anderson
localization transition.

\textcolor{black}{In the absence of magnetic fields or spin-dependent
scattering\,\cite{hikami_spin-orbit_1980}}, it has long been established\textcolor{black}{\,\cite{abrahams_scaling_1979,fisher_relation_1981}}
that the sign of loop corrections in the single-parameter scaling
equations also leads to an absence of diffusion in disordered 2D lattices.
Just as it happens in 1D, all single-particle states are immediately
localized by the presence of disorder. However, the effects in charge
transport are still qualitatively different in 1D and 2D systems despite
their apparent similarities$\,$\textcolor{black}{\cite{thouless_localization_1973,fan_linear_2021,lherbier_transport_2008}}.
In both cases, there are two different (but interrelated) length scales
that control the single-electron dynamics: \emph{i)} the mean-free
path $\p{\ell}$ which describes the long-distance decay of the \emph{disorder-averaged
single-particle propagator}\textcolor{black}{\,}\textit{\textcolor{black}{\emph{\cite{lee_disordered_1985}}}}
and \emph{ii)} the wavefunction localization length $\p{\xi}$ which
is extracted from the long-distance behaviour of the \emph{disorder-averaged
two-particle propagator}\textcolor{black}{\,\cite{janssen_fluctuations_2001}}.
In 1D these two scales are basically the same $\p{\xi\propto\ell}$$\,$\cite{thouless_localization_1973},
but in 2D systems they can get exponentially separated in the weak
disorder limit (\textcolor{black}{$\xi\propto\ell\,\exp\!\p{\pi k_{F}\ell/2}$\,\cite{lee_disordered_1985}})
driven by angular scattering channels that are absent in 1D. This
separation of scales opens up space for a sizeable crossover regime
where coherent quantum dynamics happens way below the electronic localization
length and, therefore, becomes solely dominated by the much shorter
mean-free path (i.e., $\ell<L_{\phi}<\xi$, where $L_{\phi}$ is the
electron phase coherence length). When this happens, charge transport
in a 2D system shows a diffusive character akin to that of a disordered
3D metal below its Anderson transition. Only then, electric transport
turns into a local process that makes it possible to define an intensive
conductivity - derived from the electronic quantum diffusivity via
Einstein's relation - that accurately characterizes the system's response
to an applied electric field$\,$\cite{kubo_fluctuation-dissipation_1966,markos_two-dimensional_2010,sinner_diffusive_2022}.
Despite being a well-established result, a definitive observation
of such a diffusive crossover in numerical studies of quantum transport
is a challenging problem that has been notoriously absent from the
literature. However, this matter must be addressed to guarantee that
the electrical conductivity results acquired in simulated 2D systems$\,$\cite{roche_conductivity_1997,nikolic_deconstructing_2001,cresti_charge_2008,ferreira_unified_2011,cresti_impact_2013,fan_efficient_2014,garcia_real-space_2015,garcia_kubobastin_2016,cysne_numerical_2016,joao_basis-independent_2019,wilson_disorder_2020,dlimi_crossover_2021,santos_pires_anomalous_2022}
can be appropriately understood from a physical perspective. The aim
of this work is to demonstrate that such a crossover can indeed be
directly observed in numerical studies of quantum transport in disordered
Hamiltonians.

Since the 1980s, quantum transport in solid-state models has been
studied by one of two complementary approaches: a mesoscopic approach
(the so-called Landauer-Büttiker formulation\textcolor{black}{\,\cite{landauer_spatial_1957,landauer_electrical_1970,caroli_direct_1971,mackinnon_calculation_1985,stone_what_1988}})
or a bulk response approach (Kubo’s formula and its non-linear generalizations\textcolor{black}{\,\cite{nakano_method_1956,streda_galvanomagnetic_1975,greenwood_boltzmann_1958,kubo_general_1956,bastin_quantum_1971,joao_basis-independent_2019}}).
The latter is a very general formulation of quantum transport which,
in principle, could be used to accurately describe the electrodynamic
properties of arbitrary samples independently of them being coupled
to leads or not\textcolor{black}{$\!$\cite{baranger_electrical_1989,nikolic_deconstructing_2001}}.
Additionally, it can be entirely formulated in terms of lattice Green’s
functions\textcolor{black}{$\!$\cite{streda_galvanomagnetic_1975,greenwood_boltzmann_1958,bastin_quantum_1971}.}
Their mathematical convenience sparked the development of very efficient
real-space algorithms\textcolor{black}{$\!$\cite{weisse_optical_2005,weisse_chebyshev_2004,fan_efficient_2014}}
that are able to numerically compute bulk conductivities of independent
electron models with a linear complexity in the number of orbitals.
However, despite its practicality, the bulk description of transport
has important drawbacks that become especially relevant when attempting
to capture non-local mesoscopic effects. For a start, it implicitly
assumes there is a well-defined local conductivity in the system,
something that does not hold true unless it behaves diffusively. Secondly,
when such numerical calculations are performed in finite lattices,
an effective linewidth must always be assigned to the system’s energy
levels to smooth out their discrete spectra$\,$\cite{czycholl_numerical_1980,weise_kernel_2006,fan_linear_2021,joao_kite_2020,weisse_chebyshev_2004,weisse_optical_2005}.
This technical detail means that, by design, the method is effectually
computing a space average of local conductivities that were determined
for different phase coherent regions within the sample$\,$\cite{czycholl_numerical_1980,czycholl_conductivity_1981,thouless_effect_1980}.
The space averaging of the conductivity has a crucial impact on the
results obtained within the non-local mesoscopic regime or deep in
the localized regime$\,$\cite{czycholl_conductivity_1981,thouless_conductivity_1981,mackinnon_conductivity_1980}.
Therefore, any result from bulk transport calculations is only physically
accurate if the system is in a diffusive transport regime.

In contrast, the mesoscopic approach$\,$\cite{buttiker_four-terminal_1986,landauer_spatial_1957,landauer_electrical_1970}
provides a way to analyse quantum charge transport across a sample
as a wave scattering phenomenon without any assumption of locality.
In its simplest form, the Landauer-Büttiker approach assumes that
the mesoscopic sample is connected between two electrical leads (clean
semi-infinite conductors) with the conductance of the entire sample,
$G$, being determined by the sample’s quantum transmission in the
energy band comprised by the electrochemical potentials of the two
leads$\,$\cite{stone_what_1988}. Being semi-infinite, the leads
function as free fermionic baths coupled to the sample which provide
the continuous spectrum necessary to eliminate the mean-level spacing.
Due to these two factors, the mesoscopic formulation can precisely
characterize all the system's transport regimes and makes it possible
to distinguish them from the conductance's behavior as the disorder
strength or system dimensions are changed. A localized regime, for
example, will exhibit log-normal sample-to-sample fluctuations in
the conductance, with a typical value that decreases exponentially
to zero as the distance between the two leads increases$\,$\cite{mackinnon_one-parameter_1981,santos_pires_global_2019}.
Regarding the diffusive regime, it is distinguished by a conductance
with Gaussian fluctuations and an average value that scales as $G\propto S/L$,
where $S$ is the sample's cross-sectional dimension and $L$ is the
distance between the leads$\,$\cite{mackinnon_scaling_1983}.

Unfortunately, despite its wide scope, the efficiency of mesoscopic
transport calculations poses a severe limitation on its application
for studying transport in very large systems, a particularly critical
point when trying to access 2D diffusive transport regimes. In practice,
these calculations are performed through the so-called Caroli-Mier-Wingreen
equation\,\cite{caroli_direct_1971,meir_landauer_1992}, which is
an exact Green’s function representation of the Landauer-Büttiker
formula amenable to a real-space formulation that requires the semi-infinite
leads to be imposed as non-hermitian boundary conditions on the lattice
Hamiltonian (surface-green’s functions). These boundary conditions
can be computed using iterative decimation\,\cite{sancho_highly_1985}
or eigenchannel decomposition algorithms\,\cite{mackinnon_calculation_1985,umerski_closed-form_1997,wimmer_quantum_2009}
(see Lewenkopf and Mucciolo\,\cite{lewenkopf_recursive_2013} for
a review), while the sample’s transmission coefficient is obtainable
via the real-space Recursive Green’s function method (RGFM)$\,$\cite{mackinnon_calculation_1985}.
Implementations of the RGFM are generally very memory-efficient algorithms
but feature an asymmetric scaling with the system dimensions: typically
an $\cal O\p{S^{3}\,L}$ time complexity. A more favourable (but more
memory expensive) approach is found in the Kwant transport code$\,$\cite{groth_kwant_2014,wimmer_quantum_2009},
which relies on a nested dissection algorithm\,\cite{george_nested_1973}
that shows an $\cal O\p{S^{2}\,L}$ scaling instead. In both cases,
the unfavourable non-linear scaling with the cross-sectional dimension
is a bottleneck that generally prevents the use of a mesoscopic approach
to reach and observe the diffusive crossover regime in simulated 2D
samples.

Recently, some proposals$\,$\cite{yu_chebyshev_2020,de_castro_efficient_2023,de_castro_fast_2024}
have been made on how to avoid the limitations of mesoscopic transport
approaches by means of altered spectral methods that allow electrical
contacts to be embedded on the simulated system$\,$\cite{alvermann_chebyshev_2008}.
Instead, we take on a different path and tackle the problem by a time-resolved
approach in which steady-state transport properties are determined
from the stabilized quantum dynamics of the system (coupled to truncated
leads) after being biased by an external potential. Time-resolved
approaches to quantum transport have seen an increase in popularity,
ever since their usefulness for the study of dc-transport was clearly
demonstrated in 1D systems$\,$\cite{santos_pires_landauer_2020,pal_emergence_2018}.
In addition, these techniques can also be used to study more complex
electrodynamic effects such as transient current dynamics$\,$\cite{tuovinen_distinguishing_2019,ridley_quantum_2021},
mesoscopic Bloch oscillations$\,$\cite{popescu_emergence_2017,pinho_bloch_2023}
and non-linear optical effects$\,$\cite{dimitrovski_high-order_2017,oliaei_motlagh_ultrafast_2020,silva_polaritonic_2020,rajpurohit_nonperturbative_2022,rajpurohit_ballistic_2023}
(including novel non-linear photo-galvanic effects$\,$\cite{aversa_nonlinear_1995,fregoso_jerk_2018,ventura_comment_2021,ventura_second_2020})
in simulated tight-binding systems. In this paper, we extend the methodology
from$\,$\cite{santos_pires_landauer_2020} to investigate all transport
regimes of a wide 2D nanoribbon within a two-contact transport setup.
To enable the observability of a diffusive crossover regime, we further
refine the time-dependent approach to integrate the effectiveness
of stochastic trace evaluation techniques$\,$\cite{santos_pires_landauer_2020,weisse_optical_2005}
with a bandwidth compression scheme (space-modulation of the hopping)
inside the finite leads$\,$\cite{vekic_smooth_1993}. Together, these
increments allow the accurate simulation of sufficiently wide nanoribbons
to demonstrate the 2D diffusive crossover regime.

This paper is structured as follows: Section$\,$\ref{sec:Models-and-Methods}
presents the main components of our toy-model Hamiltonian and provides
technical information on the methods used for measuring the charge
current and performing the time-evolution of the many-electron system.
The bandwidth compression scheme is described in detail in Section$\,$\ref{sec:Time-Dependent-Transport-in},
initially for 1D systems and then expanded for 2D. The main findings
of the paper are presented in Section$\,$\ref{sec:Emergence=000020Diffusive=000020Regime},
where the outcomes of our simulations clearly show a 2D diffusive
crossover. Lastly, we summarize our findings and offer a future direction
for this work in Section$\,$\ref{sec:Conclusion-and-Outlook}.

\section{Details on models and methods \protect\label{sec:Models-and-Methods}}

The aim of this work is to study time-dependent quantum transport
in a two-dimensional two-terminal system whose geometry we can control.
We follow \cite{santos_pires_landauer_2020} by splitting the system
into the sample and finite leads. Even though the leads are finite,
their properties are tailored as to accurately reproduce semi-infinite
ideal leads. Both the disorder and electric field $E$ exist exclusively
inside the sample. The electric field is uniform and is adiabatically
engaged from $t=0$, leading to a potential difference $\Delta V=EL$
across the leads (see Fig.\,(\ref{Fig:=000020nanoribbon=000020and=000020current=000020stabilization})$\,\p b$)
and giving rise to a current.

\subsection{Equilibrium Hamiltonian}

\begin{figure}[tb]
\centering{}\includegraphics[width=0.9\textwidth]{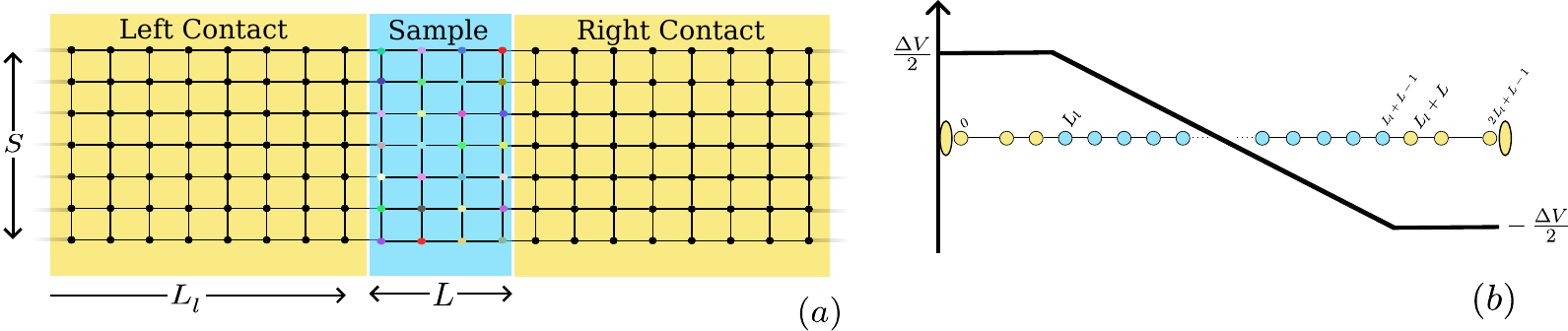}\caption{a) Geometry used in our calculations: two-dimensional tight-binding
lattice divided into disordered sample and left and right leads, with
applied potential $V\left(x\right)$. The sample's longitudinal length
is noted by L, whereas its width is described by S. Both finite sized
contacts possess a lateral length of $L_{l}$ sites. b) Longitudinal
cross-section of the spatial profile of the applied potential.}
\label{Fig:=000020nanoribbon=000020and=000020current=000020stabilization}
\end{figure}

The underlying Hamiltonian for both the sample and the leads is the
two-dimensional square lattice tight-binding Hamiltonian of hopping
parameter $w$. The sample has length $L$ and width $S$, and is
connected to the left and right leads of length $L_{l}$ and width
$S$ (see Fig.\,(\ref{Fig:=000020nanoribbon=000020and=000020current=000020stabilization})$\,\p a$).
Before the electric field is turned on, the Hamiltonian can be expressed
as a sum of five terms, 
\begin{equation}
\cal H_{0}\!=\!\mathscr{\cal H}_{\text{L}}+\cal H_{\text{R}}+\cal H_{\text{S}}+\cal V_{\text{LS}}+\cal V_{\text{SR}},
\end{equation}
where $\cal H_{\text{L}}$ ($\cal H_{\text{R}}$) stands for the Hamiltonian
of the left (right) finite contact, $\cal H_{\text{S}}$ describes
the sample, and $\cal V_{\text{SL}}$ ($\cal V_{\text{SR}}$) describes
the boundary hoppings coupling the sample to the left (right) contact.
The sample term is written as
\begin{equation}
\s{\cal H_{\t S}\,=\,}\sum_{x=L_{l}}^{L_{l}+L-1}\,\mathit{h}_{x}+\sum_{x=L_{l}}^{L_{l}+L-2}u_{x,x+1},\label{eq:=000020Hamiltonian=000020sample}
\end{equation}
where the Hamiltonian of a single slice is described as
\begin{equation}
\s{h_{x}\,=\,\sum_{y=0}^{S-1}\varepsilon\left(x,y\right)\ket{x,y}\bra{x,y}-w\sum_{y=0}^{S-2}\left[\ket{x,y+1}\bra{x,y}+\t{H.c.}\right]}.
\end{equation}
The term describing the electronic hopping, $w$, between neighbouring
slices is
\begin{equation}
\s{u_{x,x+1}\,=\,-w\sum_{y=0}^{S-1}\left[\ket{x+1,y}\bra{x,y}+\text{H.c.}\right]},
\end{equation}
where $\varepsilon(x,y)$ is an onsite disordered potential whose
values are drawn out of a box distribution in the interval $[-W/2,W/2]$
and the lattice parameter was set to unity. In order to mimic the
consequences of connecting the central device to semi-infinite leads
we employ a spatial modulation of the leads' hoppings and onsite energies
(see Section$\,$\ref{sec:Time-Dependent-Transport-in} for a detailed
explanation). For this reason, we distinguish between vertical ($t_{y}$)
and horizontal ($t_{x}$) hoppings such that the Hamiltonian of the
left lead can be written as

\vspace{-0.2cm}
\begin{equation}
\s{\cal H_{\t L}}=\sum_{x=0}^{L_{l}-1}\,h_{\t L;\,x}+\sum_{x=0}^{L_{l}-2}u_{\t L;\,x,x+1}\label{eq:Left=000020Lead=000020Hamiltonian}
\end{equation}
with 
\begin{equation}
\s{h_{\t L;\,x}\,=\,\sum_{y=0}^{S-1}U\left(x,y\right)\ket{x,y}\bra{x,y}+\sum_{y=0}^{S-2}\,t_{y}\p{x,y}\left[\ket{x,y+1}\bra{x,y}+\t{H.c.}\right]}
\end{equation}
and 
\begin{equation}
\s{u_{\t L;\,x,x+1}\,=\,\sum_{y=0}^{S-1}t_{x}\p{x,y}\left[\ket{x+1,y}\bra{x,y}+\text{H.c.}\right]}.
\end{equation}
The right contact is modelled after
\begin{equation}
\s{\cal H_{\t R}\,=\,}\sum_{x=L+L_{l}}^{2L_{l}+L-1}\,h_{\t R;\,x}+\sum_{x=L+L_{l}}^{2L_{l}+L-2}u_{\t R;\,x,x+1}
\end{equation}
with 
\begin{equation}
\s{h_{\t R;\,x}\,=\,\sum_{y=0}^{S-1}U\left(x,y\right)\ket{x,y}\bra{x,y}+\sum_{y=0}^{S-2}\,t_{y}\p{x,y}\left[\ket{x,y+1}\bra{x,y}+\t{H.c.}\right]}
\end{equation}
and 
\begin{equation}
\s{u_{\t R;\,x,x+1}\,=\,\sum_{y=0}^{S-1}t_{x}\p{x,y}\left[\ket{x+1,y}\bra{x,y}+\text{H.c.}\right]}.
\end{equation}

\noindent Finally, the connection between the three parts of the system
is given by the boundary hoppings, which are defined as

\vspace{-0.5cm}

\noindent{}
\begin{subequations}
\noindent
\begin{align}
\cal V_{\text{SR}}\! & =\!-w\sum_{y=0}^{S-1}\left[\ket{L_{l}+L,y}\bra{L_{l}+L-1,y}+\t{H.c.}\right]\\
\cal V_{LS}\! & =\!-w\sum_{y=0}^{S-1}\left[\ket{L_{l},y}\bra{L_{l}-1,y}+\t{H.c.}\right].
\end{align}
\end{subequations}
Before we describe the systematic approach that was taken to simulate
the quasi-steady state of the current's time-evolution (see Section$\,$\ref{subsec:Non-Equilibrium-Time-Evolution}
for details), it is worth noting how the mesoscopic approach referenced
in the Introduction would be applied to the study of quantum transport
within a sample described by Eq.$\,$(\ref{eq:=000020Hamiltonian=000020sample}).

\subsubsection*{The Landauer-B{\"u}ttiker Method}

\noindent For our specific context we consider the two-terminal Landauer
formula, which relates the steady-state current traversing from the
left to the right lead with the energy integral of the central sample's
quantum transmittance, $\mathcal{T}(\varepsilon)$. More precisely,
we have
\begin{equation}
I_{\text{L}\to\text{R}}^{{\scriptscriptstyle Land}}\!=\frac{e}{2\pi\hbar}\!\int_{-\infty}^{+\infty}\!\!\!d\varepsilon\left[f_{\text{F}}\left(\varepsilon+\frac{\Delta V}{2}\right)-f_{\text{F}}\left(\varepsilon-\frac{\Delta V}{2}\right)\right]\mathcal{T}(\varepsilon),\label{eq:Definition:=000020Landauer=000020Integral}
\end{equation}
where $f_{\text{F}}(x)=1/\left(1+\exp\left(\left(x-\mu\right)/k_{B}T\right)\right)$
is the Fermi-Dirac distribution. Despite the integration over energies
being performed from $-\inf$ to $+\inf$ the difference between the
Fermi-Dirac functions will severely reduce the effective range of
this computation. The integral's range is reduced to a window centered
around the Fermi energy, being controlled by the temperature, $T$
and the potential bias, $\Delta V$. From here on forward, the \emph{transmission
band }will be defined to be this particular region within the system's
spectrum. In the limit $T\rightarrow0K$, it corresponds to a window
centered around the Fermi energy, spanning from $-\Delta V/2$ up
to $\Delta V/2$. In the absence of interactions, the quantum transmittance
of the sample can be expressed within the non-equilibrium transport
formalism of Kadanoff-Baym\,\cite{kadanoff_quantum_1962} and Keldysh\,\cite{keldysh_diagram_1965},
yielding the well-known \textit{Caroli formula}\,\cite{caroli_direct_1971}:

\vspace{-0.20cm}

\begin{equation}
\mathcal{T}(\varepsilon)=\text{Tr}\left[\mathbf{G}_{\varepsilon}^{\text{a}}\cdot\boldsymbol{\Gamma}_{\varepsilon}^{\text{R}}\cdot\mathbf{G}_{\varepsilon}^{\text{r}}\cdot\boldsymbol{\Gamma}_{\varepsilon}^{\text{L}}\right],
\end{equation}
where $\mathbf{G}_{\varepsilon}^{\text{a/r}}=\left[\varepsilon\mp i0^{+}-\mathcal{H}_{S}-\Sigma_{\varepsilon}^{\text{L}}-\Sigma_{\varepsilon}^{\text{R}}\right]^{-1}$
is the sample's advanced/retarded single-particle Green's function
in the presence of both leads. Each lead dresses the single-particle
states of the sample by a self-energy, $\Sigma_{\varepsilon}^{\text{L/R}}$,
whose anti-hermitian part also defines the level-width operators $\boldsymbol{\Gamma}_{\varepsilon}^{\text{R}}$.

\subsection{Non-Equilibrium Time-Evolution\protect\label{subsec:Non-Equilibrium-Time-Evolution}}

To study the dynamics of the system in Fig.\,(\ref{Fig:=000020nanoribbon=000020and=000020current=000020stabilization})\,$\p a$,
we assume that it starts from a thermal equilibrium with temperature,
$T$ and common chemical potential, $\mu$. This is described by the
single-particle density matrix
\begin{equation}
\rho_{0}\,=\,\frac{1}{1+\exp\left[\frac{\cal H_{0}-\mu}{k_{B}T}\right]},\label{eq:FermiDirac}
\end{equation}
which is driven out of equilibrium by a potential ramp 
\begin{equation}
V(x)=\,\begin{cases}
1/2 & \qquad\;\,x<L_{l}\\
\frac{1}{L+1}\,\left(L_{l}-\frac{1}{2}+\frac{L}{2}-x\right) & \!\!\,\,\,L_{l}\leq x\leq L_{l}+L-1\\
-1/2\, & \qquad\;\,x>L_{l}+L-1
\end{cases}.
\end{equation}
such that $\Delta V\times V(x)$ gives rise to a potential difference
of $\Delta V$ between the leads as shown in Fig.\,(\ref{Fig:=000020nanoribbon=000020and=000020current=000020stabilization})\,$\left(b\right)$.
The corresponding operator for this potential bias is
\begin{equation}
\cal V=-e\!\!\!\!\!\sum_{x=L_{l}}^{L_{l}+L-1}\sum_{y=1}^{S}V(x)\ket{x,y}\bra{x,y},
\end{equation}
such that the full time-dependent Hamiltonian is

\vspace{-0.20cm}

\begin{equation}
\cal H\p t\,=\,\cal H_{0}+\Delta V\,f\p t\,\cal V,\label{eq:Definition:=000020H(t)=000020-=000020Linear-1}
\end{equation}
where the temporal profile $f\p t$ encapsulates the full time-dependence
of the perturbation. Ultimately, we want to obtain the current traversing
the sample, which requires summing up all the contributions of the
local currents along a cross-section. Let $\mathcal{I}_{x,y}$ be
the operator that represents the electrical current flowing from site
$\left(x,y\right)$ to $\left(x+1,y\right)$: 
\begin{equation}
{\displaystyle \cal I_{x,y}\,=\,\frac{ew}{i\hbar}\left(\ket{x+1,y}\bra{x,y}-\ket{x,y}\bra{x+1,y}\right)}.\label{eq:Definition:=000020Local=000020Current=000020Operator=000020Ox}
\end{equation}
The expectation value of this operator provides the current $I_{x,y}\left(t\right)$
flowing from site $\left(x,y\right)$ to $\left(x+1,y\right)$ at
time $t$ and is expressed as the following trace:

\vspace{-0.20cm}
\begin{equation}
\s{I_{x,y}\p t=\text{Tr}\left[\rho\p t\,\cal I_{x,y}\right]}\label{eq:Current=000020as=000020trace}
\end{equation}
with $\rho\p t$ being the density matrix evaluated at an arbitrary
instant. Since the scope of this paper is to compute a linear response
electrical coefficient, from this point forward we will consider a
linearized expression for the density matrix. If the perturbation
is suddenly turned on at $t=0$, the resulting current $I_{x,y}^{\text{sud}}$,
up to linear order in $\Delta V$ can be computed as

\begin{equation}
\s{I_{x,y}^{\text{sud}}\p t\,=\,\int_{0}^{t}\!\!dt^{\prime}\dot{I}_{x,y}^{\text{sud}}\p{t^{\prime}}},
\end{equation}
where the time derivative of $I_{x,y}^{\text{sud}}$ is defined as

\begin{equation}
\dot{I}_{x,y}^{\text{sud}}\p{t^{\prime}}\,=\,\frac{\Delta V}{i\hbar}\,\t{Tr}\left[\mathcal{U}_{t^{\prime}}\com{\mathcal{V}}{\rho_{0}}\,\mathcal{U}_{t^{\prime}}^{\dagger}\,\cal I_{x,y}\right],\label{eq:Transverse=000020Current=000020Explicit=000020Linear-1-1}
\end{equation}
where $\mathcal{U}_{t^{\prime}}=e^{-\frac{i}{\hbar}\mathcal{\cal H}_{0}t^{\prime}}$
is the time evolution operator of the unperturbed system from time
$0$ to $t^{\prime}$. The trace is evaluated with methods that are
based on the Kernel Polynomial Method (KPM) \cite{weisse_optical_2005,joao_kite_2020,ferreira_critical_2015}
(see Section$\,$\ref{subsec:Chebyshev-Time-Evolution-Method}). If
the profile $f\left(t\right)$ is chosen such that $f\p 0=0$ and
$f\p{t\rightarrow\inf}=1$, we will obtain a local current, 
\begin{equation}
{\displaystyle I_{x,y}\p t}=\int_{0}^{t}\!\!dt^{\prime}\dot{f}\left(t-t^{\prime}\right)I_{x,y}^{\text{sud}}\left(t^{\prime}\right),\label{eq:Transverse=000020Current=000020Explicit=000020Linear-1}
\end{equation}
whose profile is the convolution between a smoothing filter and the
time-dependent current obtained with partition-free initial conditions.
Owing to the sudden connection of the biasing potential $I_{x,y}^{\t{sud}}\p t$
oscillates along the quasi-steady state plateau. If these oscillations
are small compared with the time average of the quasi-steady state,
then this quantity agrees perfectly with the Landauer formalism prediction.
However, if the system's geometry and disorder strength place it in
the localized phase, the average value of the quasi-steady state drops
and the amplitude of the oscillations becomes comparable to it. To
address this, the uniform electric field is adiabatically connected,
with a time dependence given by $f\p t$. The measured transverse
current then results from a moving average between $I_{x,y}^{\t{sud}}\p t$,
and a kernel, $\dot{f}\p t$. Hereafter, the presented time-dependent
results were computed using $f\p t\,=\,1-e^{-\frac{t}{\tau}}$, where
$\tau$ is an adiabatic parameter.

\subsection{Chebyshev Time-Evolution Method \protect\label{subsec:Chebyshev-Time-Evolution-Method}}

While Eq.\,(\ref{eq:Transverse=000020Current=000020Explicit=000020Linear-1-1})
formally tells us how to calculate the current crossing the transverse
section of the sample, as a function of time, it still requires the
calculation of $\mathcal{U}_{t}$ and $\rho_{0}$, which are nontrivial
functions of $\mathcal{H}_{0}$. The trace is replaced by the average
over the expectation value of an ensemble of random vectors, and the
operators are replaced by a Chebyshev series expansion in powers of
the Hamiltonian. Since this series only converges on the open interval
between -1 and 1 on the real number line, the Hamiltonian has to be
shifted and rescaled such that its eigenvalues lie within this interval.
We define $\tilde{\mathcal{H}_{0}}=\left(\mathcal{H}_{0}-\lambda\right)/\Delta$
as the rescaled Hamiltonian, $\lambda$ as the shift which makes the
spectrum symmetric and $\Delta$ as the spectrum range of the Hamiltonian.
In practice, $\Delta$ is slightly larger than this to ensure the
openness of the interval. In our case, we can set $\lambda=0$ because
the spectrum is symmetrical. With this in mind, the density matrix
and the time evolution operators are expanded as

\vspace{-0.70cm}

\begin{eqnarray}
\rho_{0}\! & = & \sum_{m=0}^{\infty}\mu_{m}^{F}T_{m}\left(\tilde{\mathcal{H}}_{0}\right)\label{eq:ExpansionFermiFunction}\\
\mathcal{U}_{t} & = & \sum_{m=0}^{\infty}\mu_{m}^{U}\left(t\right)T_{m}\left(\tilde{\mathcal{H}}_{0}\right),\label{eq:=000020ExpansionTimeEvolution}
\end{eqnarray}
where $T_{m}(x)$ is a Chebyshev polynomial of the first-kind, and
the Chebyshev moments, $\mu_{m}^{F}$ and $\mu_{m}^{U}\left(t\right)$,
are defined as

\vspace{-0.60cm}
\begin{eqnarray}
\mu_{m}^{F} & = & \frac{2}{1+\delta_{m,0}}\int_{-1}^{1}\!\!dx\frac{T_{m}(x)}{\pi\sqrt{1-x^{2}}\left[1+\exp\left(\frac{x-\mu/\Delta}{k_{B}T/\Delta}\right)\right]}\\
\mu_{m}^{U}\left(t\right) & = & \frac{2}{1+\delta_{m,0}}(-i)^{m}J_{m}(w\,\Delta\,t/\hbar),
\end{eqnarray}
where $J_{m}\left(x\right)$ is a Bessel function of the first kind.
The coefficients of the Fermi function contain the full information
about the temperature and chemical potential of the initial thermal
state. In a similar way, the time dependence of the time evolution
operator is entirely captured in the $\mu_{m}^{U}\left(t\right)$
coefficients.

The trace in Eq.$\,$(\ref{eq:Transverse=000020Current=000020Explicit=000020Linear-1-1})
when evaluated in real-space can be reduced to the computation of
the following matrix elements due to the fact that the current operator
is local:

\vspace{-0.20cm}

\begin{equation}
{\displaystyle \begin{alignedat}{1}\dot{I}_{x,y}^{\text{sud}}\left(t\right) & {\displaystyle \,=\,}-2\,\Delta V\,\frac{ew}{\hbar^{2}}\,\text{Re}\,\left[\bra{x,y}\mathcal{U}_{t}\,\mathcal{V}\,\mathcal{U}_{t}^{\dagger}\,\rho_{0}\ket{x+1,y}\right.\\
 & \left.\,\,\,\,\,\,\,\,\,\,\,\,\,\,\,\,\,\,\,\,\,\,\,\,\,\,\,\,\,\,\,\,\,-\bra{x+1,y}\mathcal{U}_{t}\,\mathcal{V}\,\mathcal{U}_{t}^{\dagger}\,\rho_{0}\ket{x,y}\right].
\end{alignedat}
}\label{eq:Definition:=000020Time-Derivative=000020of=000020the=000020Current}
\end{equation}
Each of these terms can be computed by using the series expansion
of $\rho_{0}$ and $\mathcal{U}_{t}$ and the recursive properties
of the Chebyshev polynomials. More precisely, given an arbitrary state
$\ket{\psi}$, we define $\ket{\psi_{m}}$ from the action of the
Chebyshev operator $T_{m}$ on $\ket{\psi}$ and we use the recursive
properties of these polynomials to relate the different $\ket{\psi_{m}}$
among themselves, as such:

\vspace{-0.70cm}
\begin{align*}
\ket{\psi_{m}} & \equiv T_{m}\left(\tilde{\mathcal{H}}_{0}\right)\ket{\psi}=2\tilde{\mathcal{H}}_{0}\ket{\psi_{m-1}}-\ket{\psi_{m-2}}\text{ with \ensuremath{\ket{\psi_{0}}=\ket{\psi}}\text{ and }\ensuremath{\ket{\psi_{1}}}=\ensuremath{\tilde{\mathcal{H}}_{0}\ket{\psi}}}
\end{align*}
Therefore, the truncated approximations of the action of $\rho_{0}$
and $\mathcal{U}_{t}$ on $\ket{\psi}$ simply read

\vspace{-0.50cm}

\begin{align}
\rho_{0}\ket{\psi} & =\sum_{m=0}^{M_{f}-1}\mu_{m}^{F}\ket{\psi_{m}}\text{ and }\,\,\mathcal{U}_{t}\ket{\psi}=\sum_{m=0}^{M_{t}-1}\mu_{m}^{U}\left(t\right)\ket{\psi_{m}},\label{eq:ExpansionsTruncated}
\end{align}
where $M_{f}$ ($M_{t}$) is the truncation order for the expansion
of the $\rho_{0}$ ($\mathcal{U}_{t}$) operator. The calculation
of $I_{x,y}\left(t\right)$ requires the evaluation of $\dot{I}_{x,y}^{\text{sud}}\left(t\right)$
along a discrete set of times $t_{n}=n\delta t$ in the integration
interval from $0$ to $t$. This process would require computing the
time evolution operator for all times $t_{n}$. To avoid this, we
instead compute $\dot{I}_{x,y}^{\text{sud}}\left(t_{n}\right)$ in
increments of $\delta t$ reusing the objects that had already been
used for $\dot{I}_{x,y}^{\text{sud}}\left(t_{n-1}\right)$. This procedure
can be summarized as follows (see Fig.$\,$(\ref{fig:Schematics_Chebyshev_Linear})
for a diagrammatic representation):
\begin{figure}[tb]
\centering{}\includegraphics[width=0.95\textwidth]{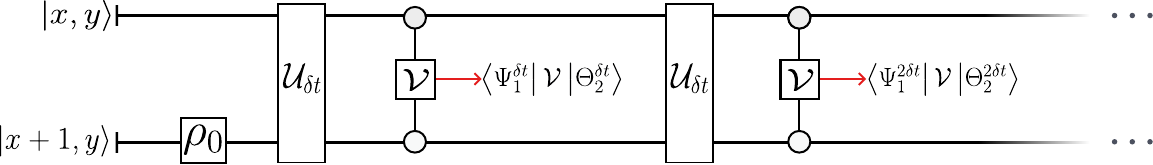}\caption{Schematics of the operations required to compute the first term of
a local current's time-derivative. \protect\label{fig:Schematics_Chebyshev_Linear}}
\end{figure}

\begin{enumerate}
\item Define the four Chebyshev vectors:
\begin{equation}
{\displaystyle \ket{\Psi_{1}^{0}}\,\equiv\,\ket{x,y},\,\,\ket{\Psi_{2}^{0}}\,\equiv\,\ket{x+1,y},\,\,\ket{\Theta_{1}^{0}}\,\equiv\,\rho_{0}\ket{x,y},\,\,\ket{\Theta_{2}^{0}}\,\equiv\,\rho_{0}\ket{x+1,y}}\label{eq:Definition:=000020Four=000020Chebyshev=000020Vectors}
\end{equation}
\item Time evolve all four vectors in a series of time-steps (sized $\delta t$):
\begin{equation}
\begin{alignedat}{1}\ket{\Psi_{1}^{t_{n+1}}} & \,=\,\mathcal{U}_{\delta t}\ket{\Psi_{1}^{t_{n}}}\\
\ket{\Psi_{2}^{t_{n+1}}} & \,=\,\mathcal{U}_{\delta t}\ket{\Psi_{2}^{t_{n}}}\\
\ket{\Theta_{1}^{t_{n+1}}} & \,=\,\mathcal{U}_{\delta t}\ket{\Theta_{1}^{t_{n}}}\\
\ket{\Theta_{2}^{t_{n+1}}} & \,=\,\mathcal{U}_{\delta t}\ket{\Theta_{2}^{t_{n}}}
\end{alignedat}
\end{equation}
\item At each time-step evaluate the time-derivative of the current as follows:
\begin{equation}
{\displaystyle \dot{I}_{x,y}^{\text{sud}}\left(t_{n}\right)\,=\,-2\Delta V\,\frac{ew}{\hbar^{2}}\,\text{Re}\,\left[\bra{\Psi_{1}^{t_{n}}}\mathcal{V}\ket{\Theta_{2}^{t_{n}}}-\bra{\Psi_{2}^{t_{n}}}\mathcal{V}\ket{\Theta_{1}^{t_{n}}}\right]}
\end{equation}
\item Repeat the procedure until time $t$ has been reached, and at the
end numerically integrate $\dot{I}_{x,y}^{\text{sud}}\left(t_{n}\right)$.
The result will be the linear current from site $\p{x,y}\rightarrow\p{x+1,y}$.
\end{enumerate}

\subsubsection{Reducing Numerical Complexity}

If the unperturbed Hamiltonian is represented as a sparse matrix in
some basis (usually in real-space), the numerical calculation of either
expression in Eq.\,(\ref{eq:ExpansionsTruncated}) involves only
sparse matrix-vector operations and, thereby, has an $\mathcal{O}\left(MN\right)$
time complexity, where $N$ is the dimension of the Hamiltonian and
$M$ is the truncation order for that operator. Note that the numerical
evaluation of $\dot{I}_{x,y}^{\text{sud}}\p t$ requires a number
of operations proportional to $N\!\times\!(TM_{t}\!+\!M_{f})$, where
$T$ is the number of points used in the time discretization. For
fixed $T$, $M_{f}$ and $M_{t}$, the time complexity is $\mathcal{O}\left(N\right)$.
The complete transverse current current requires $S$ independent
calculations, one for each $y$, totalling a time complexity of $\mathcal{O}\left(NS\right)$
and making it unfavourable for wide systems with large $S$. This
complexity can be brought down to $\mathcal{O}\left(N\right)$ with
the use of random numbers, in a similar fashion to KPM. The linear
combination of position operator eigenstates with uncorrelated random
coefficients (noted as $\ket{\xi_{x}}$) with variance one ($\overline{\xi_{y}\xi_{y^{\prime}}}=\delta_{yy^{\prime}}$),
Eq.$\,$(\ref{eq:Definition:=000020Random=000020Vectors=000020Column}),
is a core component for the stochastic measurement of the trace in
Eq.$\,$(\ref{eq:Current=000020as=000020trace})

\vspace{-0.50cm}
\begin{equation}
{\displaystyle \ket{\xi_{x}}\,\equiv\,\sum_{y=1}^{S}\xi_{y}\ket{x,y}.}\label{eq:Definition:=000020Random=000020Vectors=000020Column}
\end{equation}
Furthermore, the definition of a translation operator by a unit cell
along the longitudinal direction ($\cal T$), which acts on single-particle
position eigenstates as
\begin{equation}
{\displaystyle \cal T\ket{x,y}\,=\,\ket{x+1,y}}\label{eq:Definition:=000020Translation=000020Operator=000020->=000020ket}
\end{equation}
enables this stochastic procedure to be expressed as

\vspace{-0.20cm}

\begin{equation}
{\displaystyle I_{x}\p t\,\equiv\,\sum_{y=1}^{S}I_{x,y}\p t\,=\,\frac{2ew}{\hbar}\,\text{Im}\,\overline{\bra{\xi_{x}}\rho\p t\cal T\ket{\xi_{x}}}}.\label{eq:Definition:=000020Stochastic=000020Current}
\end{equation}
Since $\bra{\xi_{x}}\p{\dots}\ket{\xi_{x}}$ is a random number, it
has a variance associated with its distribution. To get a sufficiently
small error bar, an average across a large number $R$ of random vectors
$\ket{\xi_{x}}$ needs to be performed. For this procedure to be more
advantageous relative to summing $I_{x,y}$ over $y$, $R$ has to
be smaller than $S$$\,$\cite{joao_spectral_2023}.
\begin{figure}[tb]
\begin{centering}
\vspace{-0.20cm}
\par\end{centering}
\begin{centering}
\hspace{0.00cm}\includegraphics[width=0.85\textwidth]{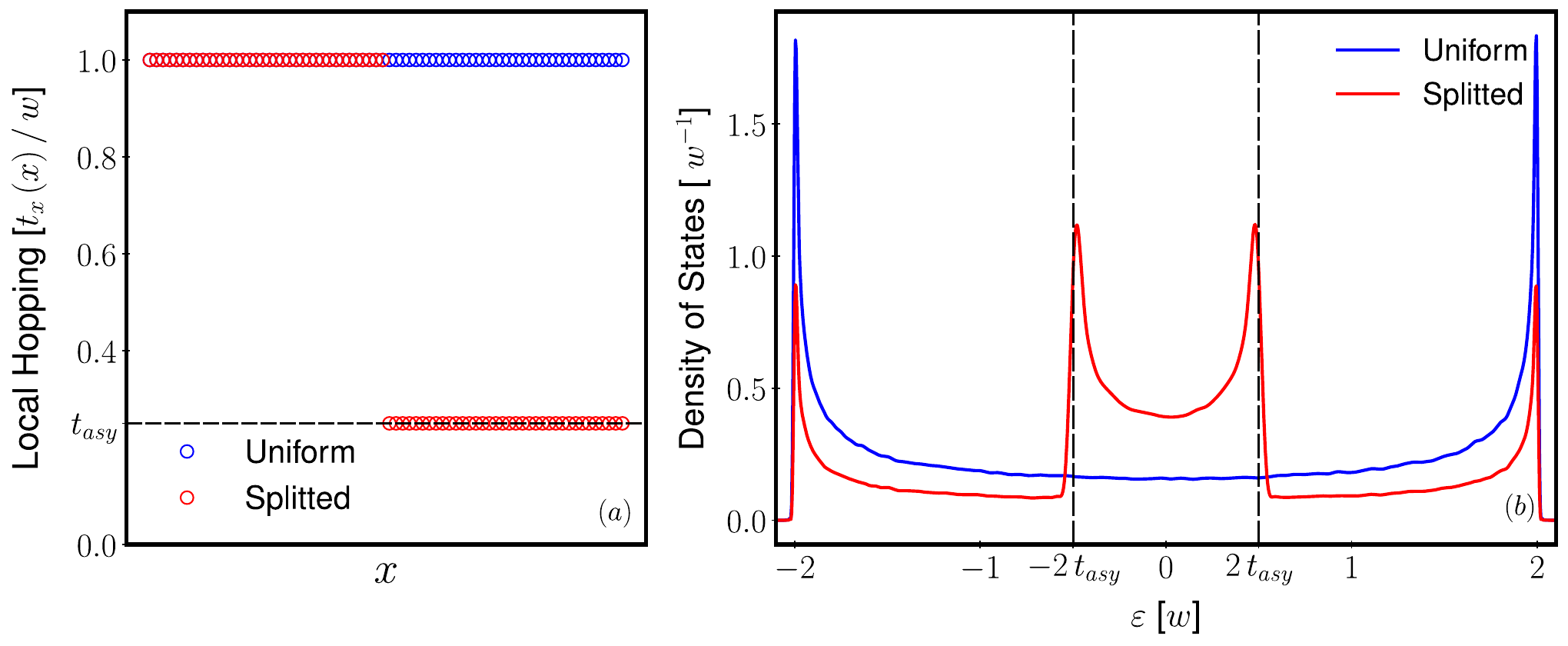}
\par\end{centering}
\begin{centering}
\vspace{0.00cm}\caption[Density of states for a 1D hopping chain with a different hopping
value at half size]{In$\,\protect\p a$ we represent the hopping term for two different
1D tight-binding chains. Whereas the blue plot corresponds to an unchanged
system, on the red curve we are altering the value of the hopping
term at half the sites of the chain. In$\,\protect\p b$ we compute
the density of states (DoS) for these two models. On one hand, we
retain the DoS for the unchanged system (with half of the total number
of states), while on the other hand we have the density of states
of a system whose hopping term is $t_{asy}$. This result shows that
we are capable of increasing the DoS on a region within the energy
spectrum that is solely controlled by $t_{asy}$.\protect\label{fig:DoS=000020Partitioned=000020System}}
\par\end{centering}
\centering{}\vspace{0.00cm}
\end{figure}

\section{Bandwidth Compression Schemes \protect\label{sec:Time-Dependent-Transport-in}}

As it was shown in$\!$\cite{santos_pires_landauer_2020}, the finite
dimension of the Hilbert space imposes constraints on the time-dependent
simulations with finite-sized leads. In particular, the duration of
the quasi-steady-state plateau ($T_{\t r}$) is limited by the reflection
time of the Fermi level states at the lead's terminations, $T_{\t r}\!=\!2\hbar L_{l}/\sqrt{w^{2}-\varepsilon_{F}^{2}}$,
with $\varepsilon_{F}$ being the system's Fermi energy. This is a
problematic feature of time-dependent simulations with finite-sized
leads since $T_{\t r}$ may prove to be too short for transients to
die out, making it impossible for the current to converge to the Landauer
quasi-steady state. An intuitive solution is to increase the size
of the finite leads, and consequentially reduce the mean-level spacing
within the transmission band, increasing $T_{\t r}$. However, due
to the number of operations being proportional to the number of Hilbert
space elements, whenever we increase the size of the leads we will
also raise the required computational effort. Therefore, we introduce
an alternative procedure that decreases the mean-level spacing within
the transmission band and, after the initial transient dies out, mimics
a de facto steady-state that would be reached on the limit of semi-infinite
leads. We will discuss how this comes about in the following sections,
building up the intuition from the 1D case.

\subsection{Reflection Time Enhancement in 1D systems\protect\label{subsec:Compression=0000201D}}

\begin{figure}[tb]
\begin{centering}
\includegraphics[width=0.9\textwidth]{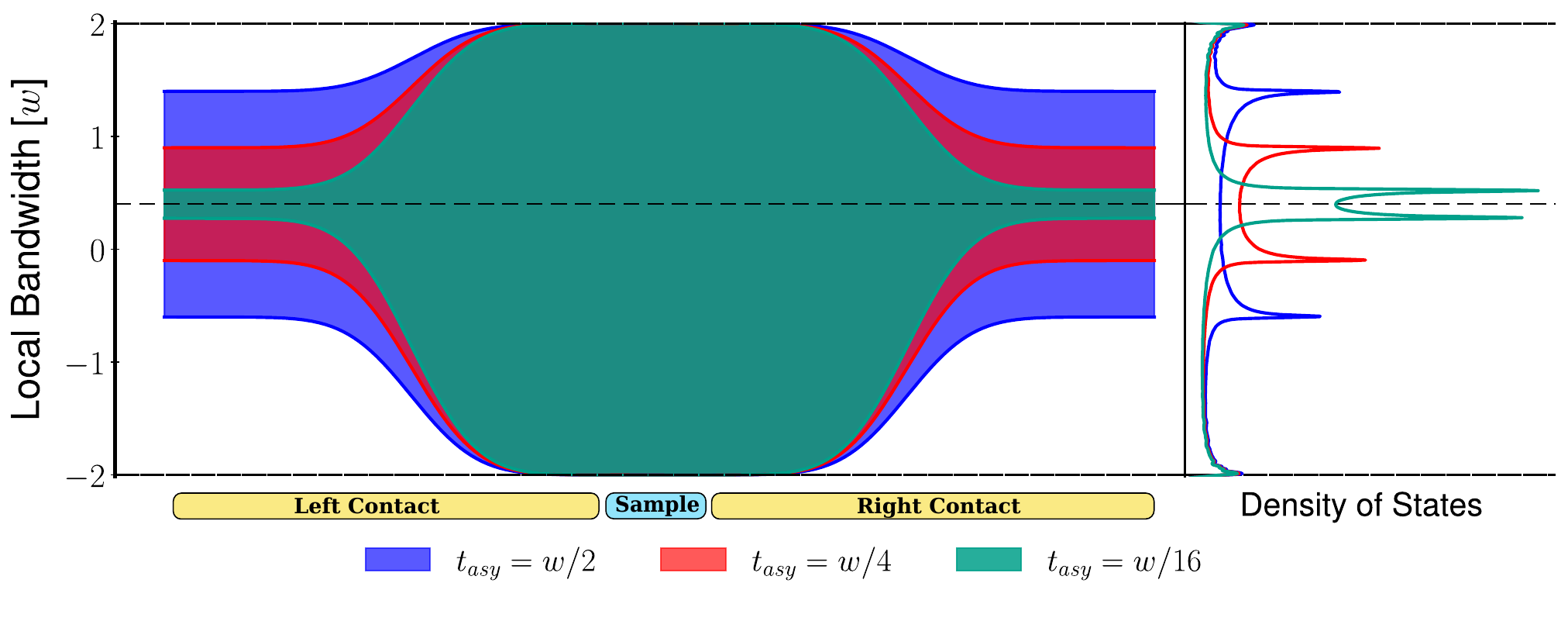}\caption{Schematics of the bandwidth compression introduced by the hopping's
modulation within a one dimensional system. The density of states
increases in the neighbourhood of the Fermi energy, and the separation
between the its maxima is approximately $4t_{asy}$, where $t_{asy}$
corresponds to the chosen asymptotic value of the hopping term. \protect\label{fig:Scheme=000020Bandwidth/DoS=000020Compression}}
\par\end{centering}
\vspace{0.00cm}
\end{figure}
\begin{figure}[tb]
\vspace{-0.20cm}
\centering{}\includegraphics[width=0.9\textwidth]{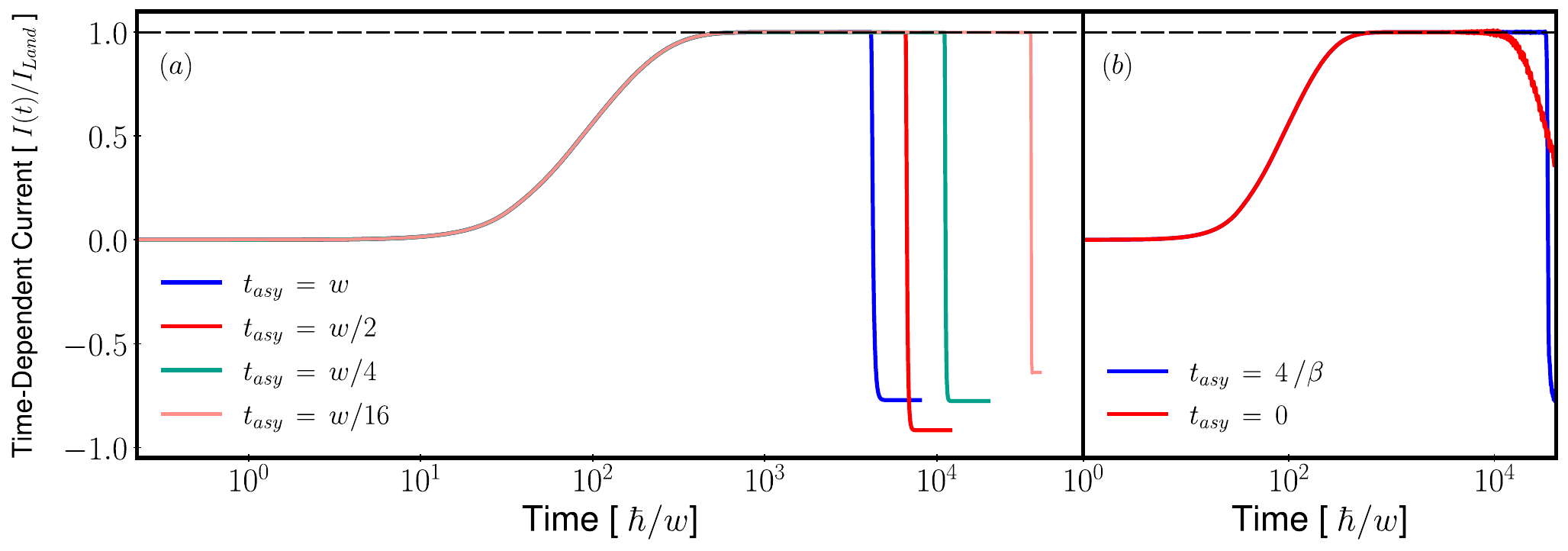}\caption{Representation of the current time-evolution for the 1D limiting case.
The simulations were performed with $\varepsilon_{F}\,=\,-0.2w,L\,=\,128,\tau\,=\,100,\beta\,=\,1024\,w^{-1}$
and $W\,=\,0.3w$ for $\protect\p a\,L_{l}\,=\,4096$ and $\protect\p b\,L_{l}\,=\,512$.It
is shown that a decrease in the asymptotic value of the hopping enables
the extension of the current's plateau. Whenever the chosen value
of $t_{asy}$ is too small, we are no longer able to ensure that all
the states covering the transmission band contribute to the maintenance
of the quasi-steady state. Consequently, in$\,\protect\p b$ an earlier
reflection is observed when $t_{asy}\,=\,0$\@. The normalization,
$I_{Land}$, was computed using the RGF method.\protect\label{fig:Currents=000020Different=000020Modulations}}
\end{figure}
As previously stated the main purpose of the introduction of a bandwidth
compression scheme is to reduce the mean-level spacing within the
transmission band. In order to motivate this procedure, we can look
at a one-dimensional toy model. Let us consider, as depicted in Fig.$\,$\ref{fig:DoS=000020Partitioned=000020System}$\,\p a$,
a tight-binding chain where the first half of the system is characterized
by a constant hopping, $t$ and the second half by $t_{\t{asy}}$.
If one computes the density of states (DoS) for this partitioned system
we will see that, although we maintain the signatures of the original
system, with local maxima at $-2w$ and $2w$, we are adding to this
DoS, the DoS of the tight-binding chain with hopping $t_{asy}$. Consequently,
this change of the hopping term leads to an increase of the DoS in
a region of the system's spectrum that is limited by $-2t_{asy}$
and $2t_{asy}$. Therefore, it is anticipated that a heuristic construction
of $t_{asy}$ allows for a reduction of the transmission band's mean-level
spacing. Complementary to this parameter, we introduce a sewing function
that should be able to glue the two different asymptotic regimes that
we are looking for. On the one hand, it has to retain the hoppings
set to unity inside the sample and on another hand, deep within the
leads' profile it should equate to $t_{asy}$ (ensuring the dense
population of the transmission band). A possible candidate for such
parametrization is
\begin{equation}
{\displaystyle f\p{x;s;\sigma}\,=\,\frac{1}{2}\,\p{\text{erf}\left[\frac{s+\sigma-x}{\sigma}\right]-\text{erf}\left[-\frac{s+\sigma+x}{\sigma}\right]}},\label{eq:Definition:=000020Factor=000020Function}
\end{equation}
where $\left\{ s,\sigma\right\} $ are two free parameters that may
be tuned to optimize the results.

The construction of the spatial modulation scheme can be attained
by directly constraining the upper limit of the local half-bandwidth,
$V_{T}$, as

\begin{equation}
V_{T}\,\equiv\,\min\p{\varepsilon_{F}+2t_{asy},2w},
\end{equation}
where the minimum with $2w$ is done in order to prevent the appearance
of states outside the original bandwidth. The lower bound for this
parametrization is similarly obtained from

\vspace{-0.30cm}

\begin{equation}
V_{B}\,\equiv\,\max\p{\varepsilon_{F}-2t_{asy},-2w},
\end{equation}
such that, at each point, the difference between $V_{T}$ and $V_{B}$
is the system's local bandwidth. The spatial dependency of the hopping
term was chosen to be 
\begin{equation}
t_{x}\p x\,=\,\frac{V_{T}\p x-V_{B}\p x}{4},\label{eq:Definition:=000020Hopping=000020Spatial=000020Profile}
\end{equation}
where
\begin{equation}
{\displaystyle \begin{alignedat}{1}V_{T}\p x & \,=\,V_{T}+\p{2w-V_{T}}\,f\p{x;s;\sigma}\\
V_{B}\p x & \,=\,V_{B}-\p{2w+V_{B}}\,f\p{x;s;\sigma}
\end{alignedat}
}.\label{eq:Definition:=000020Spatial=000020Profile=000020VT=000020and=000020VB}
\end{equation}

The sole application of Eq.$\,$(\ref{eq:Definition:=000020Hopping=000020Spatial=000020Profile})
will only increase the density of states close to zero energy. Consequently,
as one has to decrease the mean-level spacing around the system's
Fermi energy, the local potential term should also possess a spatial
modulation, that ensures the correct population of the transmission
band. Employing the spatial dependencies shown in Eq.$\,$(\ref{eq:Definition:=000020Spatial=000020Profile=000020VT=000020and=000020VB})
and considering that the lead's modulated profile should not introduce
states whose energy is not comprised within the original system's
bandwidth, one can build a modulation of the local onsite energies
as

\vspace{-0.50cm}

\begin{equation}
U\p x\,=\,\frac{V_{T}\p x+V_{B}\p x}{2},\label{eq:Definition:=000020Modulation=000020On-site=000020Energy}
\end{equation}
such that Eq.$\,$(\ref{eq:Definition:=000020Hopping=000020Spatial=000020Profile})
and Eq.$\,$(\ref{eq:Definition:=000020Modulation=000020On-site=000020Energy})
will progressively compress the total bandwidth around the Fermi energy,
as one observes in Fig.$\,$\ref{fig:Scheme=000020Bandwidth/DoS=000020Compression}.
The application of this bandwidth compression scheme is responsible
for increasing the time-dependent current's reflection time, as it
is shown in Fig.$\,$\ref{fig:Currents=000020Different=000020Modulations},
where we have represented the current's time-evolution for the same
mesoscopic setup but we consider modulation profiles whose asymptotic
values were gradually smaller. According to Fig.$\,$\ref{fig:Currents=000020Different=000020Modulations}$\,$$\p a$,
choosing smaller values of $t_{\text{asy}}$ systematically increases
the reflection time, without affecting the value of the quasi-steady-state
current. However, this improvement comes to a halt once $t_{\text{asy}}$
is so small that the asymptotic bandwidth of the modulated leads becomes
smaller than the transmission band\footnote{Due to the temperature dependence of the Fermi-Dirac distributions
a suitable choice for the hopping's asymptotic value is $t_{asy}\,=\,11k_{B}T$.}. At this point, the quasi-steady state of transport begins to decay,
as it is observed in Fig.$\,$\ref{fig:Currents=000020Different=000020Modulations}$\,\p b$.

\subsection{Extension to 2D systems}

\begin{figure}[tb]
\begin{centering}
\vspace{-0.00cm}
\par\end{centering}
\noindent\begin{raggedright}
\hspace{0.00cm}\includegraphics[width=0.94\textwidth]{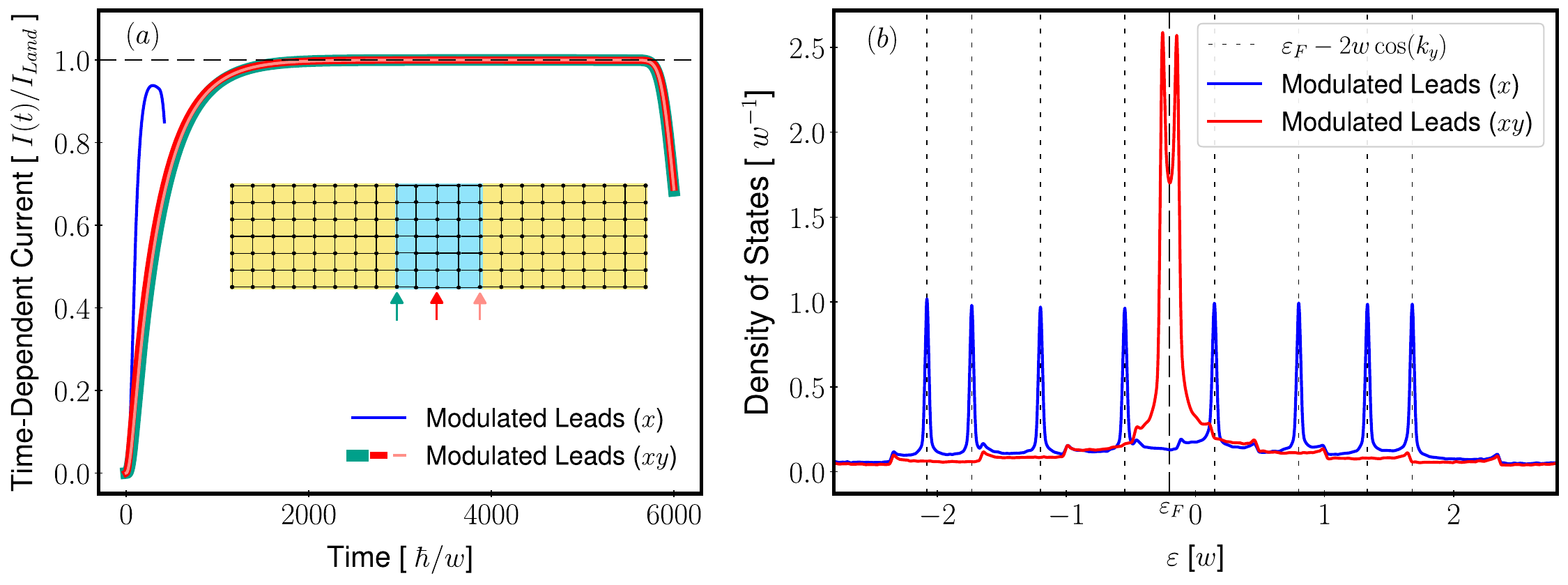}
\par\end{raggedright}
\begin{centering}
\vspace{0.00cm}\caption[Time-Evolution and density of states for a modulated 2D system]{$\protect\p a\,$Time-dependent transverse current on a disordered
nanoribbon for different simulation setups: bare leads, modulated
on the $xx$ direction and modulated on both directions. In$\,\protect\p b$
it is shown that only when $t_{y}$ is very small - deep within the
leads' profile - we are able to compress the energy levels close to
the Fermi energy. The simulations were performed with $\varepsilon_{F}\,=\,-0.2w,\,S\,=\,8,\,L\,=\,128,\,\Delta V\,=\,10^{-6}w$
and $W\,=\,0.2w$ for $L_{l}\,=\,512$. Akin to Fig.$\,$\ref{fig:Currents=000020Different=000020Modulations}
, $I_{Land}$, was evaluated using the RGF method.\protect\label{fig:Results=0000202D=000020Bandwidth=000020Compression=000020Scheme}}
\par\end{centering}
\vspace{-.20cm}
\end{figure}

Analogous to subsection$\,$\ref{subsec:Compression=0000201D} smooth
boundary conditions in 2D systems must compress the density of states
of the system within the transmission band. If one introduces the
bandwidth compression scheme leaving $t_{x}\p x$ and $U\p x$ defined
by Eq.$\,$(\ref{eq:Definition:=000020Hopping=000020Spatial=000020Profile})
and Eq.$\,$(\ref{eq:Definition:=000020Modulation=000020On-site=000020Energy}),
while keeping the vertical hoppings, $t_{y}$, unchanged and set to
$w$ we would not be able to resolve the current's quasi-steady state.
This statement is verified in the blue and green plots of Fig.$\,$\ref{fig:Results=0000202D=000020Bandwidth=000020Compression=000020Scheme}$\p a$.
Complementary to this information, in Fig.$\,$\ref{fig:Results=0000202D=000020Bandwidth=000020Compression=000020Scheme}$\,\p b$
we show that are not able to increase the density of states on the
transmission band. Instead each local maxima is centered around $\varepsilon_{F}-2w\cos k_{y}$,
where $k_{y}$ are the allowed wave-numbers of a hard-wall tight-binding
chain. In order to understand why this is happening we may look at
the Hamiltonian of a clean system. Since the spatial dependence of
the longitudinal hoppings is independent of the $y$ coordinate, we
can factorize this Hamiltonian as 
\begin{equation}
\s{\cal H\,=\,\sum_{k_{y}}\cal H^{\t{1D}}\p{k_{y}}\tensorproduct\ket{k_{y}}\bra{k_{y}}},
\end{equation}
where
\begin{equation}
\cal H^{\t{1D}}\p{k_{y}}\,=\,\s{\sum_{x}t_{x}\p x\ket{x+1}\bra x+\t{H.c.}+\sum_{x}\left[U\p x-2w\cos k_{y}\right]\ket x\bra x}.\label{eq:Effective=000020Hamiltonian=000020ky}
\end{equation}
 Therefore, the introduction of smooth boundary conditions along the
longitudinal hoppings alone, will compress each strip of the nanoribbon
in an independent manner, controlled by the diagonal term in Eq.$\,$(\ref{eq:Effective=000020Hamiltonian=000020ky}).
The motivation towards the solution that we have implemented is best
realized by restricting the possible functional forms of the vertical
hoppings, $t_{y}\p{x,y}$ to be functions of $x$ alone, $t_{y}\p{x,y}\rightarrow t_{y}\p x$,
while maintaining $t_{x}\p x$ and $U\p x$ defined by Eq.$\,$(\ref{eq:Definition:=000020Hopping=000020Spatial=000020Profile})
and Eq.$\,$(\ref{eq:Definition:=000020Modulation=000020On-site=000020Energy}).
Additionally, we considered that $t_{y}\p x$ should be constructed
from Eq.$\,$(\ref{eq:Definition:=000020Hopping=000020Spatial=000020Profile}),
whose asymptotic value, $t_{y,asy}$, should be set as a small constant\footnote{We chose to fix it to $t_{asy}/16$. Setting $t_{y,asy}=0$ would
imply that $k_{y}$ stopped being a good quantum number at the beginning
of a lead, since the Fermi surface for an eigenstate of energy, $\varepsilon$,
would be solely defined by $k_{x}$, $2t_{asy}\cos k_{x}=\varepsilon$.}. The red plot in Fig.$\,$\ref{fig:Results=0000202D=000020Bandwidth=000020Compression=000020Scheme}$\,\p a$
and$\,\p b$ confirms that this prescription is able to extract the
Landauer conductance from a time-resolved approach in 2D systems and
increase the density of states in the transmission band. 

To justify the physical grounds for this heuristic solution we begin
by looking at an exact eigenstate of the modulated lead, $\ket{\psi_{\alpha}}$,
with energy $\varepsilon_{\alpha}$. The projection of such state
onto a quantum state for which $k_{y}$ is a good quantum number,
$\ket{\lambda_{k_{y}}}$ , will be a sinusoidal function

\vspace{-0.30cm}

\begin{equation}
\s{\psi_{\alpha,k_{y}}\p x\,\equiv\,\braket{\lambda_{k_{y}}}{\psi_{\alpha}}\,=\,A\sin\p{k_{x}x+\phi}},\label{eq:Ansatz=000020Harmonic=000020Solution-1}
\end{equation}
where $A$ is the wavefunction's amplitude, $\phi$ is its phase and
$k_{x}$ is the allowed wave-number along the longitudinal direction.
The bandwidth compression scheme for 2D systems can be seen as a way
to have all (or most) eigenstates of the system with finite leads
concentrated in energies that cover the transmission band, while having
$\left\{ k_{x},k_{y}\right\} $ locally well defined near the sample,
where the hopping is constant. Therefore it is expected that the projected
eigenstate to have a well-defined $k_{x}$ in that region and its
value is such that it occupies the Fermi surface. Although the presence
of modulation does not permit the definition of $k_{x}$ as a good
quantum number, we still aim at defining a local value for it using
the following procedure.
\begin{figure}[t]
\raggedright
\begin{centering}
\vspace{-0.20cm}
\par\end{centering}
\begin{centering}
\hspace{0.00cm}\includegraphics[width=0.9\textwidth]{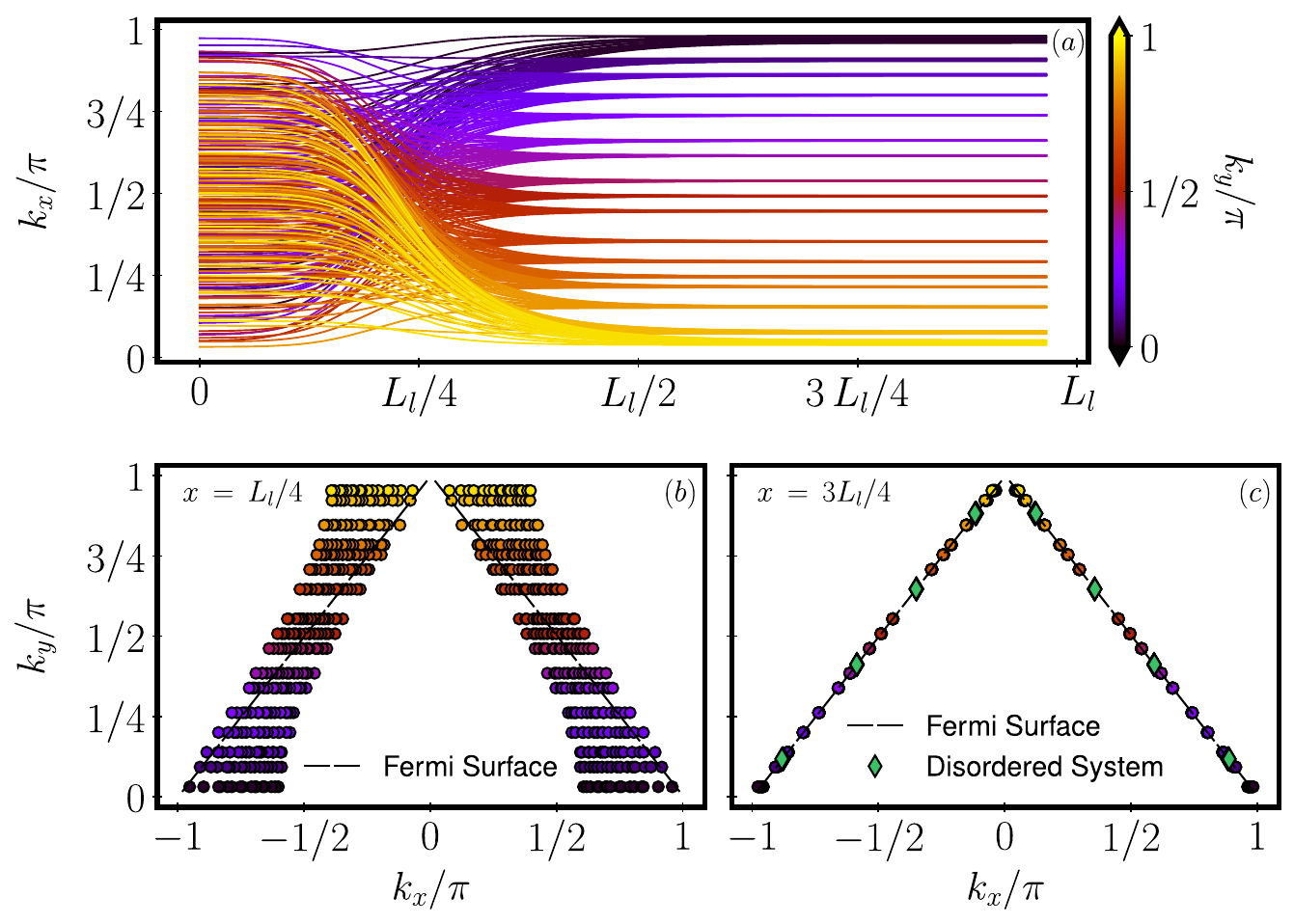}
\par\end{centering}
\begin{centering}
\vspace{-0.50cm}\caption[Representation of the population of the Fermi surface, as one introduces
smooth boundary conditions in 2D systems]{$\!\!\,\protect\p a\,$Representation of $k_{x}\protect\p x$ for
a set of fixed transverse momenta, $k_{y}$. In$\,\protect\p b$ and$\,\protect\p c$
we observe cross-sectional cuts of$\,\protect\p a$ at $x=L_{l}/4$
and $x\,=\,3L_{l}/4$. The green data sets shown in$\,\protect\p c$
verify that this approach is also valid for the case in which the
leads are connected to a disordered sample. All panels were computed
with $L_{L}\,=\,512$, whereas the disordered data was evaluated with
$S\,=\,16$ and $W\,=\,1.5w$.\protect\label{fig:Schematics=000020Fermi=000020Surface=0000202D}}
\par\end{centering}
\vspace{-0.20cm}
\end{figure}

The propagation of a projected eigenstate from the beginning of the
left lead up to the sample can be expressed through the transfer matrix,
$\mathbb{T}_{x}^{k_{y}}$, as
\begin{equation}
\s{\left[\begin{array}{c}
\psi_{x+1}\\
\psi_{x}
\end{array}\right]\,=\,\mathbb{T}_{x}^{k_{y}}\,\left[\begin{array}{c}
\psi_{x}\\
\psi_{x-1}
\end{array}\right]}
\end{equation}
 with

\vspace{-0.80cm}

\begin{align}
\s{\mathbb{T}_{x}^{k_{y}}\,=\,\left[\begin{array}{cc}
\frac{\varepsilon_{F}-2t_{y}\p x\cos k_{y}-U\p x}{t_{x}\p x} & -\frac{t_{x}\p{x-1}}{t_{x}\p x}\\
1 & 0
\end{array}\right]} & .\label{eq:Transfer=000020Matrix=000020-=000020x}
\end{align}
Therefore, knowing the projected wave-function in two adjacent sites,
$x$ and $x+1$, we could extend the projected eigenstate, by the
multiplication of the transfer matrix associated to each lattice site.
In fact, if one wishes to obtain the wave-function, $M$ sites to
the front, it is equivalent to the repeated application of the transfer
matrix
\begin{equation}
\s{\s{\left[\begin{array}{c}
\psi_{x+M}\\
\psi_{x-1+M}
\end{array}\right]\,=\,}\prod_{i=0}^{M-1}\mathbb{T}_{x+i}^{k_{y}}\,\left[\begin{array}{c}
\psi_{x}\\
\psi_{x-1}
\end{array}\right]}
\end{equation}
To define a local $k_{x}\p x$ we construct an extended state from
the successive multiplication by the matrices, $\mathbb{T}_{x}$,
\begin{equation}
\s{\left[\begin{array}{c}
\psi_{x+M}\\
\psi_{x-1+M}
\end{array}\right]\,=\,}\left[\mathbb{T}_{x}^{k_{y}}\right]^{M}\,\left[\begin{array}{c}
\psi_{x}\\
\psi_{x-1}
\end{array}\right]
\end{equation}
Since these matrices have constant hoppings they generate semi-infinite
states that are harmonic functions and have a well-defined $k_{x}$.
This will be our definition of $k_{x}$: it is the one generated by
the local transfer matrix, $\mathbb{T}_{x}^{k_{y}}$.

In Fig.$\,$\ref{fig:Schematics=000020Fermi=000020Surface=0000202D}$\,\p a$
we show the spatial dependency of $k_{x}$, as one constructs and
fits these extended states at each lattice site. The different realizations
of the colour palette marks a choice of $k_{y}$ and as we have previously
stated, within the regions in which the hoppings and onsite potential
are severely compressed, we do not have a correlation between $k_{x}$
and $k_{y}$. This characteristic is evident by the lack of coherence
between the distribution of colours for the regions of the plot that
are to the left of $L_{l}/4$. The progression along the profile of
the lead brings about a smooth change of $k_{x}$ that terminates
on a constant value. This is precisely the one permitted by the condition
that each pair, $\left\{ k_{x},k_{y}\right\} $, belongs to the Fermi
surface of the infinite lead, as one gets closer to the central device.
This conclusion is also pinpointed at Fig.$\,$\ref{fig:Schematics=000020Fermi=000020Surface=0000202D}$\p b$
and$\,\p c$ where we see cross-sectional cuts of panel$\,\p a$.
In$\,\p b$ we show that the pairs $\left\{ k_{x},k_{y}\right\} $
do not lie within the Fermi surface. Contrastingly, in$\,\p c$ we
are sufficiently close to the sample and each value of $k_{x}$ that
was previously scattered along the Brillouin zone is now correctly
mapped towards the Fermi surface.

This was not only verified for a modulated lead but also for the full
system with modulated leads connected to a disordered sample. Taking
the eigenstates within the transmission band and projecting them on
$k_{y}$, we get $k_{x}$ values that are on the Fermi surface, as
seen in the green diamonds of Fig.$\,$\ref{fig:Schematics=000020Fermi=000020Surface=0000202D}$\p c$.

\section{Emergence of a Diffusive Transport Regime\protect\label{sec:Emergence=000020Diffusive=000020Regime}}

Using the setup described earlier, we analyse the conductance, $G\p{L,S}$
of disordered systems as a function of their longitudinal length,
$L$, and width, $S$, at a fixed Anderson disorder strength, $W=0.9w$.
The time-resolved approach described in Section$\,$\ref{sec:Time-Dependent-Transport-in}
is crucial to measure the conductance for larger systems, $S\geq2048$,
whereas an implementation of the RGF method was employed for systems
with smaller cross-sections. In the diffusive regime the conductance
should scale as $G=\sigma S/L$, where $\sigma$ is the system's conductivity.
Therefore, by studying the scaled conductance, $g\,\equiv\,GL/S$,
one can unmistakably identify the system's diffusive behaviour: it
occurs when $g$, as a function of $L$ for fixed $S$, remains constant,
matching the conductivity.

\begin{figure}[t]
\begin{centering}
\includegraphics[width=0.9\textwidth]{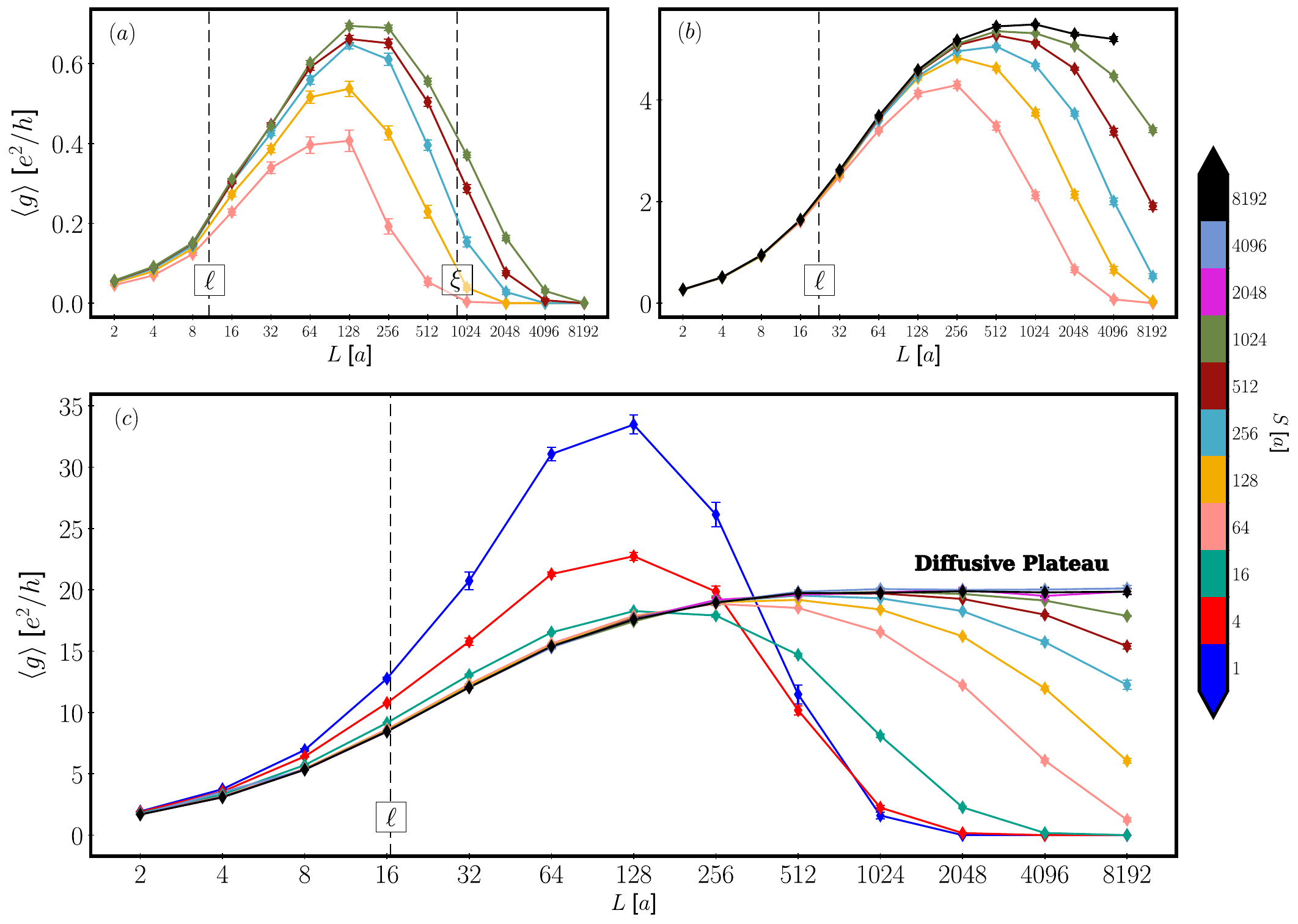}
\par\end{centering}
\caption{Disordered-averaged scaled conductance as a function of the mesoscopic
devices geometries, for a fixed Anderson disorder strength, $W=0.9w$.
In$\,\protect\p a$ we have fixed $\varepsilon_{F}=-3.985w$, whereas
in$\,\protect\p b$ we have $\varepsilon_{F}=-3.8w$ and in$\,\protect\p c$
$\varepsilon_{F}=0.0w$. Only at half-filling, the mean-free path
and localization length scales are sufficiently far apart, so that
we are able to observe the formation of a diffusive plateau. \protect\label{Fig:=000020diffusion=000020graph}}
\end{figure}

Each curve in Fig.\ref{Fig:=000020diffusion=000020graph} shows the
rescaled conductance's scaling behaviour with the device length, for
fixed $S$. Additionally, each panel features vertical dashed lines
indicating estimations of both the mean-free path, $\ell$, and localization
length, $\xi$. The latter is only displayed in$\,\p a$ because,
only for this particular set of parameters, $\xi$ falls within the
range of the simulated device lengths. The mean-free path was estimated
by studying the disorder-averaged Green's function, $G^{\t r}$, of
a large periodic disorder sample, with identical hoppings and disorder
strength to Eq.$\,$(\ref{eq:=000020Hamiltonian=000020sample}). This
quantity was obtained from the relation between the disordered-averaged
Green's function and its clean counterpart$\!$\cite{altland_condensed_2010},
$G_{\t{cl}}^{\t r}$,
\begin{equation}
\s{\left|\av{G^{\t r}\p{\bb x,\bb x^{\prime},\varepsilon_{F}}}\right|\,=\,\left|G_{\t{cl}}^{\t r}\p{\bb x,\bb x^{\prime},\varepsilon_{F}}\right|e^{-\frac{\left|\mathbf{x}-\mathbf{x}^{\prime}\right|}{2\ell}}},\label{eq:Disorder=000020Green's=000020Functions}
\end{equation}
where $\av{\dots}$ represents disorder averaging and $G^{\t r}\p{\bb x,\bb x^{\prime},\varepsilon_{F}}=\bra{\bb x}G^{\t r}\p{\varepsilon_{F}}\ket{\bb x^{\prime}}$.
We have approximated the measurement of the mean-free path in the
thermodynamic limit by employing twisted boundary conditions in both
directions. By doing so, $G^{\t r}$ was averaged over different $512\times512$
disordered supercells, while $G_{\t{cl}}^{\t r}$ was solely averaged
over different twist-angle configurations. Thereafter, the mean-free
path was computed by fitting the exponential decay shown in Eq.$\,$(\ref{eq:Disorder=000020Green's=000020Functions}).
Whenever charge transport is compatible with the ballistic regime
$\av g$ becomes exponential in $\log L$. This dependency is observed
throughout the plots in Fig.$\,$\ref{Fig:=000020diffusion=000020graph},
for $L<\ell$. The fitted values of $\ell$ mark the end of the ballistic
behaviour and for $L>\ell$ the ballistic-diffusive crossover begins.

The estimation of the localization length can be directly made from
the conductance scaling behaviour, since for a localized system, $G\sim S\exp\p{-2L/\xi}$.
 The vertical line shown in Fig.\ref{Fig:=000020diffusion=000020graph}$\,\p a$
was extracted from the normalized conductance curve with the highest
cross-section, $S\!=\!1024$. It is shown that due to the closeness
between the mean-free path and localization length scales we are not
able to observe the formation of a diffusive plateau. From this panel
to$\,\p b$ we move from $\varepsilon_{F}=-3.985w$ to $\varepsilon_{F}=-3.8w$,
which increases the distance between $\ell$ and $\xi$. Despite this,
the separation between these scales is still not sufficient for a
sizeable diffusive regime to be observed. We further note that the
plot with $S=8192$ doesn't span all values of $L$. Currently the
simulation of non-equilibrium currents on localized samples is still
at an early stage of development. We have observed that the initial
transient associated with the non-equilibrium current significantly
increases as one moves into the localized regime. This particular
challenge will be the focus of future publication.

The localization length was drastically increased in Fig.$\,$\ref{Fig:=000020diffusion=000020graph}$\p c$
where we moved to the half-filling case, $\varepsilon_{F}=0.0w$.
For small $S$, we find quasi one-dimensional behaviour. As $L$ is
increased, the conductance decreases slower than $1/L$, causing an
initial increase in the disordered-averaged normalized conductance,
$\av g$. For sufficiently large $L$, a localized regime takes over
and $\av g$ drops to zero. In quasi one-dimensional systems where
$S<\ell<\xi$, it is commonly accepted that the proximity between
the mean-free path and localization length scales hinders the development
of the diffusive regime. As S increases, the mean free path grows;
simultaneously, the localization length also increases at a much faster
pace due to the exponential dependence. This separation manifests
itself in the broadening of the maximum of $\av g$. For sufficiently
large $S$ ($S\gtrsim512$), several things happen:
\begin{enumerate}
\item $\xi$ becomes large enough that the localized regime is unobservable
within the range of parameters we were able to simulate.
\item As $\ell$ converges to the two-dimensional value with increasing
$S$, all the curves collapse into the same curve.
\item A diffusive regime develops for $L\gtrsim2048$, where $\av g$ is
constant and equal to the conductivity $\sigma$.
\end{enumerate}
\begin{figure}[tb]
\begin{centering}
\includegraphics[width=0.9\textwidth]{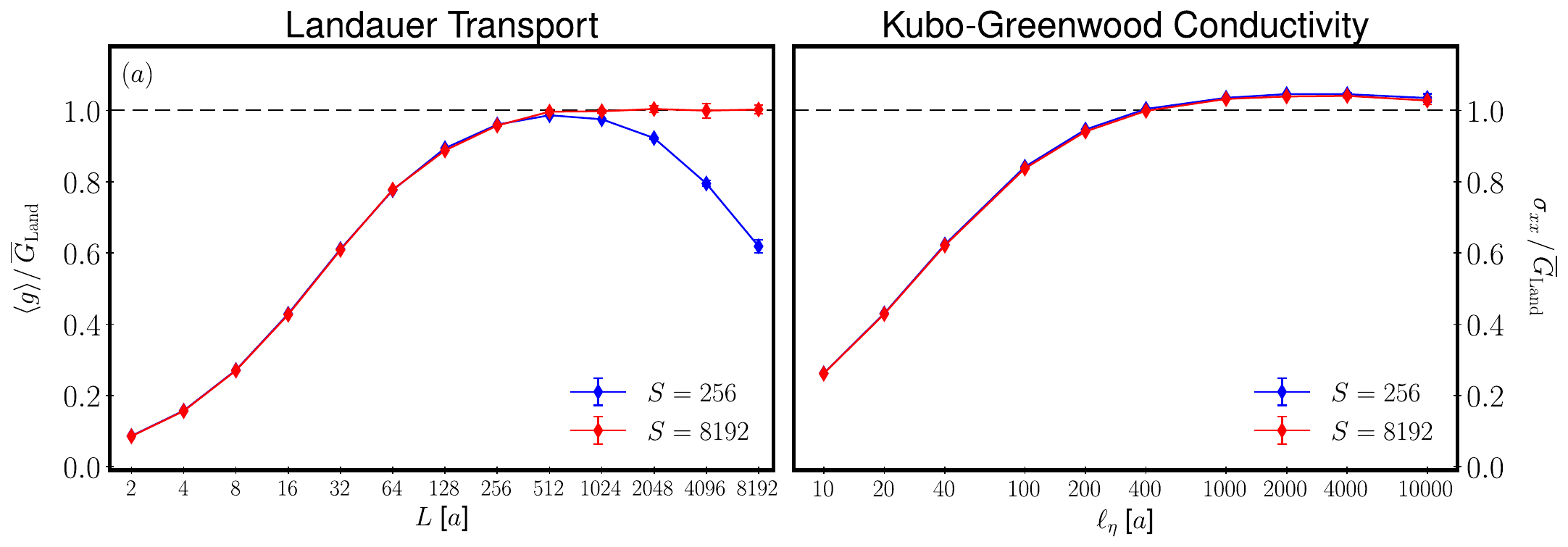}
\par\end{centering}
\caption{Direct comparison between the scaling behaviour, at half-filling,
between the conductance$\,\protect\p a$ and the Kubo-Greenwood conductivity$\,\protect\p b$.
The plots are normalized by the extracted Landauer conductivity, $\overline{G}_{\protect\t{Land}}$.}
\label{Fig:=000020kubo=000020comparasion}
\end{figure}

Another common method to determine the  zero temperature longitudinal
conductivity of a sample is through perturbation theory using the
Kubo-Greenwood formula$\,$\cite{van_tuan_anomalous_2016,fan_linear_2021}
in a large periodic and disordered system without leads: 
\begin{equation}
\sigma_{xx}(E)=\frac{e^{2}\pi\hbar}{\Omega}\t{Tr}[\delta(E-\hat{H})\hat{V_{x}}\delta(E-\hat{H})\hat{V_{x}}],\label{eq:Kubo-Greenwood=000020formula}
\end{equation}
where $\hat{V}_{x}$ is the velocity operator in the longitudinal
direction. To establish a close connection to our time-dependent results,
we chose a system with exactly the same geometry, hopping and disorder
parameters as the sample in Eq.$\,$(\ref{eq:=000020Hamiltonian=000020sample}).
The transverse direction has open boundary conditions, whereas the
longitudinal direction has twisted boundary conditions. 

The numerical method described in this manuscript is based on quantum
transmission and consistently describes both local and nonlocal transport
regimes. In contrast, Eq.$\,$(\ref{eq:Kubo-Greenwood=000020formula})
assumes from the start that transport is local, i.e. diffusive, where
a bulk conductivity is properly defined. The numerical evaluation
of Eq.$\,$(\ref{eq:Kubo-Greenwood=000020formula}) for finite systems
relies on introducing a phenomenological inelastic parameter, $\eta$,
that corresponds to the resolution of a single Dirac-Delta. The results
obtained within this framework are highly dependent on the value acquired
by $\eta$$\,$\cite{nikolic_deconstructing_2001}, and the common
procedure is to study the conductivity as a function of this parameter$\,$\cite{fan_linear_2021}.
To obtain physical results, $\eta$ should be larger than the finite
system's mean-level spacing, and because this formulation scales linearly
with the geometry of the system$\,$\cite{ferreira_critical_2015,braun_numerical_2014}
it is a very powerful tool from a numerical standpoint.

In Fig.$\,$\ref{Fig:=000020kubo=000020comparasion}$\,\p a$ we reprised
two plots from Fig.$\,$\ref{Fig:=000020diffusion=000020graph}, while
in$\,\p b$ we represented the Kubo-Greenwood conductivity's scaling
with the spectral resolution, $\eta$, computed with KITE \cite{joao_kite_2020}.
When this quantity is determined using Eq.$\,$(\ref{eq:Kubo-Greenwood=000020formula}),
we do not have access to a concrete length scale. It is common to
define an effective inelastic length scale, $\ell_{\eta}$, as $\ell_{\eta}=\hbar v_{F}/\eta$,
where $v_{F}$ is the Fermi velocity. Despite the underpinning assumptions of the Kubo approach to quantum
transport, Eq.\,(\ref{eq:Kubo-Greenwood=000020formula}) is broadly
expected to yield the same results as any transmission-based approach
whenever the system's response to the applied electric fields becomes
local or, equivalently, if charge transport happens diffusively across
the system. This comparison is made explicit in Fig.$\,$\ref{Fig:=000020kubo=000020comparasion},
where it is shown that, while the $\sigma_{xx}$ Kubo conductivity
initially grows with $\ell_{\eta}$ (resembling a ballistic transport
regime), it eventually settles into a plateau for intermediate values
of this scale suggesting a diffusive behaviour with a size-independent
conductivity. Even though the plateau of the Kubo conductivity bares
striking similarities to the diffusive plateaus obtained earlier for
the Landauer conductance of the two-terminal devices, it is worth
remarking that the two values do not perfectly agree, showing an approximately
4\% relative mismatch in this particular setup. This minute difference
can be attributed to the fact that the diffusive behaviour in two-dimensional
systems only exists as a crossover regime, as any disordered two-dimensional
electron gas is ultimately localized in the absence of external magnetic
fields \cite{abrahams_scaling_1979}. Hence, the assumption of local
transport in a two-dimensional nanorribon is only approximately true.

As a final remark, the comparison in Fig.$\,$\ref{Fig:=000020kubo=000020comparasion}
further highlights the very different way in which localization affects
the transport results within a Kubo or Landauer approach. The plots
present numerical results obtained for nanorribons having two different
cross-sections, $S=256$ and $S=8192$, corresponding to localization
lengths that  differ by orders of magnitude. In the two-terminal setup,
Fig.$\,$\ref{Fig:=000020kubo=000020comparasion}$\p a$, we see a
clear difference in the rescaled conductance for $L>512$, revealing
the two quantum transport regimes: the localized $\left(S=256\right)$
and the diffusive $\left(S=8192\right)$. From the Kubo-Greenwood
results, Fig.$\,$\ref{Fig:=000020kubo=000020comparasion}$\p b$,
both geometries present an undistinguishable behaviour, exposing the
difficulties of the Kubo-Greenwood formalism to capture localization
effects in mesoscopic systems.

\section{Conclusion and Outlook\protect\label{sec:Conclusion-and-Outlook}}

In this paper, we obtained the conductance of a disordered sample
connected to leads, in a two-dimensional tight-binding square lattice,
as a function of its geometry. The methods developed in this work
enabled us to accurately obtain the conductance of samples with cross-sections
ranging from one to several thousand unit cells, allowing a clear
separation of scales between the localization length and the mean
free path. Thus, a diffusive regime is observable across a wide range
of geometries and a two-dimensional conductivity can be defined. Our
method shows perfect agreement with the Landauer formula and very
good agreement with the Kubo-Greenwood formula in the diffusive regime.
Owing to the long transients that emerge in wide localized devices,
the application of this technique to the precise measurement of conductance
in systems over the diffusive-localized crossover has not yet been
realized.

Our method explicitly includes the leads in the Hilbert space of the
simulation and captures the temporal profile of the current as an
electric field is applied adiabatically inside the sample. The conductance
is extracted from the developed non-equilibrium quasi-steady state.
In order to extend this quasi-steady state and allow for a more accurate
reading without any additional computational effort, a spatial modulation
is applied to the leads' hoppings that closely mimics the semi-infinite
lead limit. The expectation value of the current is obtained by resorting
to a stochastic evaluation of the trace, replacing a sum over local
currents by an average over expectation values of random vectors on
cross-sectional currents. The numerical complexity thus becomes linear
in the cross-sectional width, rather than quadratic, and the trade-off
between the error bar and numerical complexity proves beneficial for
wide (width $>1000$ unit cells) samples. The time evolution of the
system is performed via a Chebyshev expansion of the density matrix
and the time evolution operators, leveraging the sparsity of the real-space
Hamiltonian while being numerically exact and scaling linearly with
the size of the Hilbert space both in time and space complexity.

This new approach is the result of several developments in real-space
simulation methods and offers a clear picture of transport in two-dimensional
disordered systems. It is easily generalizable to more complex models
and geometries. Additionally it can also be used to compute several
other quantities, such as local charge and spin density, as time-dependent
quantities in the presence of both resonant and non-resonant scatterers.
An immediate extension of the present work is the study of the different
transport regimes in disordered nanowires, in which a metal-to-insulator
phase transition and a stable diffusive metal phase are known to exist
\cite{schubert_distribution_2010}. In the  diffusive metal phase
the conductivity obtained from the conductance's scaling behaviour
should exactly match the one obtained using Kubo formalism. Even though
we have focused exclusively on the non-equilibrium quasi-steady state,
the transient regime is also accessible and should provide valuable
information about the systems being studied. 

\section*{Acknowledgments}

This work was supported by Fundação para a Ciência e a Tecnologia
(FCT, Portugal) in the framework of the Strategic Funding UIDB/04650/2020.
Further support from Fundação para a Ciência e a Tecnologia (FCT,
Portugal) through Projects No.EXPL/FISMAC/0953/2021 (J.M.A.P.), 2022.15885.CPCA,
2023.11029.CPCA and Grant. No. 2023.02155.BD (J.M.A.P) are acknowledged.
S.M.J. further acknowledges funding from the Royal Society through
a Royal Society University Research Fellowship URF\textbackslash R\textbackslash 191004
and funding from the EPSRC programme grant EP/W017075/1.

\bibliographystyle{SciPost_bibstyle}
\bibliography{References}

\end{document}